\documentclass[prd,amsmath,twocolumn,10pt,superscriptaddress,floatfix,nofootinbib]{revtex4-2}

\usepackage[utf8]{inputenc}
\usepackage{amsmath,amssymb,amsfonts,bm}
\usepackage{epstopdf}
\usepackage{graphicx}
\usepackage{xcolor}
\usepackage{bbm}
\usepackage{multirow}
\usepackage{easyReview}
\usepackage{mathtools}
\usepackage{dsfont}
\usepackage{inputenc}
\usepackage{gensymb}
\usepackage{appendix}
\usepackage[
colorlinks=true,
linkcolor=magenta,
breaklinks=true,
urlcolor=blue,
citecolor=blue]{hyperref}

\newcommand{\ucas}{\affiliation{School of Physical Sciences, University of Chinese Academy of Sciences, Beijing 100049, China}}

\begin{document}
	
	\title{Algorithms for partial wave amplitudes under covariant $L$-$S$ scheme}
	\author{Hao-Jie Jing}
	\email{jinghaojie@ucas.ac.cn}
	\ucas
	\author{Shu-Ming Wu}
	\email{wushuming@ucas.ac.cn}
	\ucas
	\author{Jia-Jun Wu}
	\email{wujiajun@ucas.ac.cn}
	\ucas
	\affiliation{Southern Center for Nuclear-Science Theory (SCNT), Institute of Modern Physics, Chinese Academy of Sciences, Huizhou 516000, China}
	
	\date{\today}

	\begin{abstract}
		
		With the continuous accumulation of data from hadron collision experiments, efficient Partial Wave Analysis tools are indispensable for constructing a clear hadron spectrum. 
		%
			Currently, automated computations of scattering amplitudes primarily use the helicity scheme and covariant effective Lagrangian method.
		The automated calculations under the covariant $L$-$S$ scheme, which is one of the commonly used partial wave analysis schemes, have not been fully realized.
	    In this work, we provide a general algorithm for computing partial wave amplitudes under the covariant $L$-$S$ scheme. This will lay the foundation for automated computation of partial wave amplitudes under the covariant $L$-$S$ scheme.
		
	\end{abstract}
	
	\maketitle
	
	\section{Introduction}
	\label{sec:intro}
	
	Quantum Chromodynamics (QCD) is the cornerstone theory of strong interactions, elucidating the dynamics between quarks and gluons. 
	At low energies, quarks and gluons are confined within color-neutral hadrons due to the strong color force. 
	However, directly describing hadron formation from QCD at these energy scales is challenging due to the breakdown of perturbative field theory and the absence of nonperturbative techniques. 
	As a result, effective field theory methods have been devised, leveraging hadrons as the fundamental degrees of freedom to explore their structure and spectra, offering insights into strong interactions at low energies. 
	
	Experimentally, properties of hadronic resonances are typically inferred through two primary avenues of information. 
	%
		Firstly, resonance masses and lifetimes are deduced from line shapes in invariant mass distributions of final-state particles, known as invariant mass spectrum analysis. 
	Additionally, the spin and parity of hadronic resonances can be gleaned from angular distributions of final-state particles using Partial Wave Analysis (PWA). 
	Integration of information from both methods allows for constructing the hadron spectrum, paving the way for elucidating hadron structure.
	
	PWA is a standard approach for extracting quantum numbers of hadronic resonances. 
	It projects scattering amplitudes into partial wave amplitudes with determined total angular momentum $J$ and parity $P$ or determined orbital angular momentum $L$ and total spin $S$ quantum numbers.

	There are various commonly employed schemes for PWA, including early helicity schemes~\cite{doi:10.1142/9789814280389_0007,Jacob:1959at}, Zemach formalism~\cite{Zemach:1965ycj}, and subsequently developed schemes, which exhibit manifest Lorentz covariance\footnote{If replacing the four-momenta in the amplitude with those from another frame is sufficient to obtain the corresponding amplitude in that frame, then we say that this amplitude exhibits manifest Lorentz covariance.}, 
	such as covariant helicity schemes~\cite{PhysRevD.48.1225}, covariant effective Lagrangian method~\cite{Nakayama:2002mu,Benmerrouche:1996ij,Liang:2002tk}, and covariant orbital-spin ($L$-$S$) coupling schemes~\cite{Zou:2002ar,Zou:2002yy,Dulat:2005in,Dulat:2011rn}. 

	Compared to early PWA schemes, the schemes with manifest Lorentz covariance making them more convenient to various cascade decay processes. 
	Moreover, the fixed $L$-$S$ quantum numbers in covariant $L$-$S$ scheme help distinguish contributions from different partial wave amplitudes and facilitate the introduction of form factors dependent on $L$. 
	The scheme was initially proposed in references~\cite{Zou:2002ar, Zou:2002yy} for constructing partial wave amplitudes of $\psi MM$ and $N^*NM$, and later applied to radiative decay processes such as $N^*N \gamma$ and $\psi M \gamma$~\cite{Dulat:2005in, Dulat:2011rn}, serving as one of the main PWA schemes adopted by the BESIII Collaboration~\cite{BESIII:2023wfi, BESIII:2023sbq, BESIII:2023xac, BESIII:2023htx}. 
	
	In previous works~\cite{Zou:2002ar, Zou:2002yy, Dulat:2005in, Dulat:2011rn}, general formulas for partial wave amplitudes containing arbitrary spin particles were not provided, except for the case of vector meson decays (automatic calculation of such partial wave amplitudes has been realized in the GPUPWA package~\cite{Berger:2010zza,Liu:2014pea,gpupwa}) and the case of excited-state nucleons decay to ground-state nucleons. 
	Thus, it is challenging to achieve automatic calculation of partial wave amplitudes for arbitrary processes under the covariant $L$-$S$ scheme. 
	Automatic calculation of partial wave amplitudes has been realized under helicity scheme and covariant effective Lagrangian method, such as the TF-PWA package~\cite{tf-pwa} and the FDC-PWA package~\cite{fdc-pwa}.

	Recently, a comprehensive theoretical framework has been introduced for constructing partial wave amplitudes with manifest Lorentz covariance, encompassing particles with arbitrary spin and mass~\cite{Jing:2023rwz}. 
	Drawing upon irreducible tensors of the Lorentz group and its little groups, the authors outlined general procedures for constructing Lorentz covariant partial wave amplitudes. 
	In fact, these procedures are also applicable for constructing partial wave amplitudes in schemes without manifest Lorentz covariance. 
	In this paper, within the framework, we provide concise formulas for numerically computing arbitrary Lorentz covariant partial wave amplitudes, as well as various cases including massless particles and identical particles. 
	This lays the foundation for the subsequent implementation of automatic partial wave amplitude calculations for various processes under the covariant $L$-$S$ coupling scheme.

	This paper is organized as follows: in Sec.~\ref{sec:PWF}, we introduce the partial wave formula (PWF) of 2-body decay (2BD) process; in Sec.~\ref{sec:manybodydecay}, we cover the PWFs of cascade decay processes; in Sec.~\ref{sec:identical}, we discuss the PWFs involving identical particles; the Sec.~\ref{sec:summary} provides a summary.

	\section{PWF for 2BD}
	\label{sec:PWF}
	
	In Ref.~\cite{Jing:2023rwz}, the authors introduced a method for decomposing scattering amplitudes into partial wave amplitudes. 
	In this method, one first defines the ``pure-spin part" and ``pure-orbital part" within the amplitude, and then utilize the irreducible tensors of the SU(2) group to decompose each part separately, resulting in partial wave amplitudes with definite $L$-$S$ quantum numbers. 
	Consequently, different definitions of the ``pure-spin part" and ``pure-orbital part" will yield different partial wave amplitudes. For amplitudes of 2BD ($1\to 2+3$), the authors adopted the following definitions:
	\begin{align}
		\left[\mathcal{M}_{1}^*\right]_{\sigma_1}^{\sigma_2 \sigma_3}=~&\underbrace{\Gamma_{\alpha_1}^{\beta_2 \beta_3}\left(k_1,p_2^*,p_3^*\right) D_{\beta_2}^{~~\alpha_2}\left(\Lambda^*_2\right) D_{\beta_3}^{~~\alpha_3}\left(\Lambda^*_3\right)}_{\text {pure-orbital part}}\notag\\
		&\times \underbrace{\bar{u}_{\sigma_1}^{\alpha_1}\left(s_1\right) u_{\alpha_2}^{\sigma_2}\left(s_2\right) u_{\alpha_3}^{\sigma_3}\left(s_3\right)}_{\text {pure-spin part}},
		\label{eq:def_pure_LS_H}
	\end{align}
	where the symbol ${\cal M}$ denotes any amplitude and the superscript ``*" denotes a quantity in the standard momentum
	\footnote{For massive particles, the standard momentum is taken as the rest momentum $k_\mu=(m,0,0,0)_\mu$; for massless particles, the standard momentum is taken as the momentum along the $z$-axis $k_\mu=(|\mathbf{k}|,0,0,|\mathbf{k}|)_\mu$ with a given magnitude $|\mathbf{k}|$.} 
	frame (s.m. frame) of particle-1; $\Gamma_{\alpha_1}^{\beta_2 \beta_3}\left(k_1,p_2^*,p_3^*\right)$ denotes the coupling structure and its arguments are four-momenta of the initial and final state particles; 
	$s_i$ and $\sigma_i~(i=1,2,3)$ are the spin and $z$-axis polarization component of the particle-$i$, respectively; 
	$u_{\alpha_i}^{\sigma_i}\left(s_i\right)/\bar{u}_{\sigma_i}^{\alpha_i}\left(s_i\right)~(i=1,2,3)$ are covariant/contravariant spin wave functions in the s.m. frame of the particle-$i$, respectively; 
	$\Lambda^*_i(i=2,3)$ denote Lorentz transformations from the s.m. frame of particle-$i$ to the s.m. frame of particle-1, respectively, with $u_{\beta_i}^{\sigma_i}\left(p^*_i,s_i\right)=D_{\beta_i}^{~~\alpha_i}\left(\Lambda^*_i\right)u_{\alpha_i}^{\sigma_i}\left(s_i\right)$. 
	The rationale behind this definition is that the angular dependence in the amplitude is entirely contributed by the ``pure-orbital part"; 
	hence, the angular dependence terms in the coupling structure and spin wave functions should both be considered as part of the ``pure-orbital part".
	
	In contrast, there is another definition can be adopted:
	\begin{align}
		\left[\mathcal{M}_{2}^*\right]_{\sigma_1}^{\sigma_2 \sigma_3}=~&\underbrace{\Gamma_{\alpha_1}^{\alpha_2 \alpha_3}\left(k_1,p_2^*,p_3^*\right)}_{\text {pure-orbital part}}\notag\\
		&\times \underbrace{\bar{u}_{\sigma_1}^{\alpha_1}\left(s_1\right) u_{\alpha_2}^{\sigma_2}\left(p_2^*,s_2\right) u_{\alpha_3}^{\sigma_3}\left(p_3^*,s_3\right)}_{\text {pure-spin part}}.
		\label{eq:def_pure_LS_C}
	\end{align}
	The rationale behind this definition is that the angular dependence in the spin wave functions belongs to the ``pure-spin part", so the ``pure-orbital part" only contains the coupling structure. 
	The partial wave amplitudes provided by these two definitions are as follows:
	\begin{align}
		\left[\mathcal{H}^*_{L,S}\right]_{\sigma_1}^{\sigma_2 \sigma_3}\equiv~& \Gamma_{\alpha_1}^{\alpha_2 \alpha_3}\left(k_1,p_2^*,p_3^*;L,S\right)\notag\\
		&\times\bar{u}_{\sigma_1}^{\alpha_1}\left(s_1\right) u_{\alpha_2}^{\sigma_2}\left(s_2\right) u_{\alpha_3}^{\sigma_3}\left(s_3\right),	\label{eq:def_cov_pwa_LS_cmf_H}\\
		\left[\mathcal{C}^*_{L,S}\right]_{\sigma_1}^{\sigma_2 \sigma_3}\equiv~& \Gamma_{\alpha_1}^{\alpha_2 \alpha_3}\left(k_1,p_2^*,p_3^*;L,S\right)\notag\\
		&\times\bar{u}_{\sigma_1}^{\alpha_1}\left(s_1\right) u_{\alpha_2}^{\sigma_2}\left(p_2^*,s_2\right) u_{\alpha_3}^{\sigma_3}\left(p_3^*,s_3\right),	\label{eq:def_cov_pwa_LS_cmf_C}
	\end{align}
	where the symbol ${\cal H}_{L,S}$ and ${\cal C}_{L,S}$ denote the partial wave amplitude obtained based on Eq.~\eqref{eq:def_pure_LS_H} and Eq.~\eqref{eq:def_pure_LS_C}, respectively;  
	$\Gamma_{\alpha_1}^{\alpha_2 \alpha_3}\left(k_1,p_2^*,p_3^*;L,S\right)$ denotes the coupling structure with specific partial wave components $L$-$S$, the specific form of which can be found in Eq.~(2.6) in Ref.~\cite{Jing:2023rwz}.
	It can be demonstrated that the partial wave amplitudes within Eq.~\eqref{eq:def_cov_pwa_LS_cmf_H} are consistent with those under the helicity scheme (as in Ref.~\cite{Jacob:1959at}), 
	while the partial wave amplitudes within Eq.~\eqref{eq:def_cov_pwa_LS_cmf_C} are consistent with those under the covariant $L$-$S$ scheme (as in Ref.~\cite{Zou:2002ar}). 
	The differences between the partial wave amplitudes under these two schemes have been discussed in some works~\cite{Filippini:1995yc,Chung:1997jn}. 
	It is straightforward that the angular distributions obtained from Eq.~\eqref{eq:def_cov_pwa_LS_cmf_H} and Eq.~\eqref{eq:def_cov_pwa_LS_cmf_C} are not identical for the same $L$-$S$ quantum numbers. 
	As mentioned above, the key point is the distinct definitions of the ``pure-orbital part" and ``pure-spin part", which essentially involve choosing different bases to expanding the scattering amplitude. 
	For specific process, as long as all possible combinations of $L$-$S$ are considered, the two schemes are equivalent.
	
	While Eq.~\eqref{eq:def_cov_pwa_LS_cmf_H} and Eq.~\eqref{eq:def_cov_pwa_LS_cmf_C} are sufficient for constructing PWF for any 2BD process, these equations involve many Lorentz covariant indices, and the contraction between these indices entail considerable computational complexity, especially for processes involving particles with higher-spin. 
	Therefore, to improve the efficiency of numerical calculations, we need to further simplify Eq.~\eqref{eq:def_cov_pwa_LS_cmf_H} and Eq.~\eqref{eq:def_cov_pwa_LS_cmf_C} so that they can meet the requirements for efficient PWA of experimental data. 
	In the following subsections, we will introduce the PWF for numerical calculations.

	\subsection{2BD PWF in the s.m. frame}
	\label{sec:PWF_cmf}
	
	We first consider the partial wave amplitudes in Eq.~\eqref{eq:def_cov_pwa_LS_cmf_H}. After further simplification, one can obtain:
	\begin{equation}
		\begin{split}
			\left[{\cal H}^*_{L,S}\right]_{\sigma_1}^{\sigma_2\sigma_3}=~&\frac{|\mathbf{p}_2^*|^L}{\sqrt{2s_1+1}}\left(C_{s_1}^{SL}\right)_{\sigma_1}^{\sigma_S\sigma_L}\left(C^{s_2s_3}_{S}\right)_{\sigma_S}^{\sigma_2\sigma_3}\times\\&{\cal Y}_{L,\sigma_L}\left(\hat{\mathbf{p}}^{*}_{2}\right),
		\end{split}
		\label{eq:PWF_A_cmf}
	\end{equation}
	where $\mathbf{p}_i^*~(i=2,3)$ is the three-momentum of particle-$i$ in the s.m. frame of particle-1;  
	$\sigma_L=\sigma_1-\sigma_S$ and $\sigma_S=\sigma_2+\sigma_3$ are the $z$-axis polarization components of orbital angular momentum $L$ and total spin $S$, respectively; 
	${\cal Y}_{L,\sigma_L}\left(\hat{\mathbf{p}}\right)$ is the normalized spherical harmonic. 
	Clearly, the angular dependence is entirely given by spherical harmonics. 
	Indeed, the form of Eq.~\eqref{eq:PWF_A_cmf} is entirely consistent with the PWF under the helicity scheme~\cite{Jacob:1959at}.

	Furthermore, we consider the partial wave amplitudes in Eq.~\eqref{eq:def_cov_pwa_LS_cmf_C}. 
	After further simplification, the corresponding PWF can be obtained by rescaling the polarization indices in Eq.~\eqref{eq:PWF_A_cmf} as follows:\footnote{To obtain this result, we construct partial wave amplitude under the covariant $L$-$S$ scheme and employed Rarita-Schwinger spin wave functions~\cite{Rarita:1941mf}. 
		Interested readers may refer to Ref.~\cite{Jing:2023rwz}.}
	\begin{equation}
		\left[{\cal C}^*_{L,S}\right]_{\sigma_1}^{\sigma_2\sigma_3}=\sum_{\sigma_2',\sigma_3'}\left[{\cal H}^*_{L,S}\right]_{\sigma_1}^{\sigma_2'\sigma_3'}{\cal I}^{(s_2)\sigma_2}_{\sigma_2'}\left(\mathbf{p}_2^*\right){\cal I}^{(s_3)\sigma_3}_{\sigma_3'}\left(\mathbf{p}_3^*\right).
		\label{eq:PWF_B_cmf}
	\end{equation}
	The explicit form of ${\cal I}^{(s)\sigma}_{\sigma'}\left(\mathbf{p}\right)$, which is a dilation around the direction of momentum $\hat{\mathbf{p}}$ , is
	\begin{equation}
		{\cal I}^{(s)\sigma}_{\sigma'}\left(\mathbf{p}\right)=\sum_{\sigma''}D^{(s)\sigma''}_{\sigma'}\left(R_{\hat{\mathbf{p}}}\right)I\left(|\mathbf{p}|,s,\sigma''\right)D^{(s)\sigma}_{\sigma''}\left(R^{-1}_{\hat{\mathbf{p}}}\right),
	\end{equation}
	where $D^{(s)\sigma}_{\sigma'}\left(R\right)$ is the Wigner-$D$ matrix; $R_{\hat{\mathbf{p}}}$ is a rotation that rotates the $z$-axis in $\hat{\mathbf{p}}$; 
	$I\left(|\mathbf{p}|,s,\sigma''\right)$ is the scale of the dilation and the specific form is shown in the Appx.~\ref{appd:scale}. 
	It can be seen that the rescaling transformation provides additional angular dependence on PWF.

	\subsection{2BD  PWF in any frame}
	\label{sec:PWF_any_frame}
	
	Although the PWF provided in Eqs.~\eqref{eq:PWF_A_cmf} and \eqref{eq:PWF_B_cmf} are simpler compared to Eqs.~\eqref{eq:def_cov_pwa_LS_cmf_H} and \eqref{eq:def_cov_pwa_LS_cmf_C}, they do not exhibit manifest Lorentz covariance. 
	Therefore, it is necessary to specify the transformation rules for these PWF under Lorentz transformations. 
	By investigating the transformation behavior of scattering amplitudes that possess manifest Lorentz covariance, one can derive the transformation rules for Eqs.~\eqref{eq:PWF_A_cmf} and \eqref{eq:PWF_B_cmf}. 
	In conclusion, the scattering amplitudes in any frame can be transformed into those in the s.m. frame through the alignment rotation\footnote{The alignment rotation is also known as the Wigner-Thomas rotation, see e.g. Sec.~6.7 in Ref.~\cite{2013srgf.book.....G}.}. 
	For 2BD, the PWF in any frame (with particle momenta denoted as $\mathbf{p}_1,\mathbf{p}_2,\mathbf{p}_3$) can be expressed as follows:
	\begin{equation}
		\left[{\cal A}_{L,S}\right]_{\sigma_1}^{\sigma_2\sigma_3}=\sum_{\sigma_2',\sigma_3'}\left[{\cal A}^*_{L,S}\right]_{\sigma_1}^{\sigma_2'\sigma_3'}D^{(s_2)\sigma_2}_{\sigma_2'}\left(R_{2}\right)D^{(s_3)\sigma_3}_{\sigma_3'}\left(R_{3}\right).
		\label{eq:PWF_any_frame}
	\end{equation}
	The alignment rotation $R_{i}=R_{\hat{\mathbf{n}}_i}\cdot R_z\left(\psi_i\right)\cdot R^{-1}_{\hat{\mathbf{n}}_i}~(i=2,3)$, where $R_z\left(\psi_i\right)$ is a rotation around the $z$-axis by an angle $\psi_i$. 
	The angle $\psi_i$ and the direction $\hat{\mathbf{n}}_i$ are given by the following equation\footnote{The function $\tan^{-1}[x, y]=-i\log\left(\frac{x+ i y}{\sqrt{x^2+y^2}}\right)$. 
		When $x^2+y^2=1$, it yields the angle $\varphi \in [-\pi,\pi)$ such that $x = \cos\varphi$ and $y = \sin\varphi$.}:
	\begin{equation}
		\begin{split}
			\frac{\psi_i}{2}&=\tan^{-1}{\left[\sqrt{\frac{(E_1+m_1)(E_i+m_i)}{(E_1-m_1)(E_i-m_i)}}+\cos\theta_{1i},\sin\theta_{1i}\right]},\\
			\hat{\mathbf{n}}_i&=\frac{\hat{\mathbf{p}}_1 \times \hat{\mathbf{p}}_i}{\sin\theta_{1i}},
		\end{split}
		\label{eq:massive_wignerd}
	\end{equation}
	where $\cos\theta_{1i}=\hat{\mathbf{p}}_1 \cdot \hat{\mathbf{p}}_i~(i=2,3)$; $m_i$ and $E_i=\sqrt{m_i^2+|\mathbf{p}_i|^2}~(i=1,2,3)$ is the mass and energy of the particle-$i$, respectively.
	
	By replacing ${\cal A}^*_{L,S}$ in Eq.~\eqref{eq:PWF_any_frame} with ${\cal H}^*_{L,S}~({\cal C}^*_{L,S})$ from Eq.~\eqref{eq:PWF_A_cmf}~(Eq.~\eqref{eq:PWF_B_cmf}), one can obtain the PWF in any frame under the helicity scheme (covariant $L$-$S$ scheme), denoted as PWF-H (PWF-C).

	\subsection{2BD PWF with massless particles}
	\label{sec:massless}
	
	The PWF-H~(C) introduced in the previous section can only applies to cases with massive particles. 
	However, only minor changes need to be made for cases including massless particles.\footnote{The difference between massive and massless particles arises from the fact that their four-momenta correspond to different little groups~\cite{Bargmann:1948ck} of the Lorentz group.}
	
	At first, one needs to replace the $z$-axis polarization component $\sigma$ with helicity $\lambda$ in PWF for massless particles. 
	For instance, when both final state particles are massless, we have the following PWF in the s.m. frame:
	\begin{equation}
		\left[{\cal A}^*_{L,S}\right]_{\sigma_1}^{\lambda_2\lambda_3}\hspace{-0.6mm}=\hspace{-0.8mm}\sum_{\sigma_2,\sigma_3}\hspace{-0.8mm}\left[{\cal A}^*_{L,S}\right]_{\sigma_1}^{\sigma_2\sigma_3}\hspace{-0.5mm}D^{(s_2)\lambda_2}_{\sigma_2}\hspace{-0.8mm}\left(R_{\hat{\mathbf{p}}^*_2}\right)\hspace{-0.5mm}D^{(s_3)\lambda_3}_{\sigma_3}\hspace{-0.8mm}\left(R_{\hat{\mathbf{p}}^*_3}\right),
		\label{eq:PWF_M_cmf}
	\end{equation}
	where the symbol $\cal{A}$ can be either $\cal{H}$ or $\cal{C}$, corresponding to PWF-H or PWF-C.
	
	In addition, Eq.~\eqref{eq:PWF_M_cmf} only holds in the s.m. frame. 
	One needs the alignment rotation to obtain PWF in any frame. 
	Since the helicity of a massless particle is a Lorentz scalar, one can expect the PWF in any frame is as follows:
	\begin{equation}
		\left[{\cal A}_{L,S}\right]_{\sigma_1}^{\lambda_2\lambda_3}=\sum_{\lambda_2',\lambda_3'}\left[{\cal A}^*_{L,S}\right]_{\sigma_1}^{\lambda_2'\lambda_3'}D^{(s_2)\lambda_2}_{\lambda_2'}(\tilde{R}_{2})D^{(s_3)\lambda_3}_{\lambda_3'}(\tilde{R}_{3}),
		\label{eq:PWF_M}
	\end{equation}
	where $\tilde{R}_{i}=R_z\left(\psi_i\right)~(i=2,3)$ is a rotation around the direction $\hat{z}\equiv(0,0,1)$ by an angle $\psi_i$.
	The explicit form of the rotation parameter is $\psi_i=\psi^+_i+\psi^-_i$ with
	\begin{align}
		\psi_i^{\pm}=\tan^{-1} &\Bigg[\hat{\mathbf{p}}_1\cdot\hat{\mathbf{p}}_i \mp \hat{\mathbf{p}}_1 \cdot \hat{z}+\sqrt{\frac{(E_1+m_1)}{(E_1-m_1)}}\left(1 \mp \hat{\mathbf{p}}_i \cdot \hat{z}\right),\notag\\
		&~~\pm \left(\hat{\mathbf{p}}_1 \times \hat{\mathbf{p}}_i\right) \cdot \hat{z}\Bigg].
		\label{eq:massless_trans_angle}
	\end{align}

	Combining Eq.~\eqref{eq:PWF_M_cmf} and Eq.~\eqref{eq:PWF_M}, one can derive the PWF in any reference frame containing massless particles. 
	It is worth noting that, since massless particles have only transverse polarization components, the number of linear independent bases in the partial wave amplitudes is less than the total number of possible $L$-$S$ combinations. 
	To avoid introducing redundant bases, one needs to select a set of linear independent and complete bases from possible $L$-$S$ combinations. 
	Indeed, weight functions introduced in Ref.~\cite{Jing:2023rwz} (see Appx.~G) can be used to select a set of linear independent and complete bases.

	\section{PWF of cascade decay}
	\label{sec:manybodydecay}
	
	The previous sections introduced the PWF for 2BD. For a multi-body cascading decay, one can construct the PWF by using the 2BD PWF recursively. 
	For example, considering a 3-body decay with $1\to 2+X$ and $X\to 3+4$, the corresponding PWF can be written as follows:
	\begin{align}
		\left[{\cal A}_{L_1,S_1,L_2,S_2}\right]_{\sigma_1}^{\sigma_2\sigma_3\sigma_4}=\sum_{\sigma_X}&\left(~\left[{\cal A}_{L_1,S_1}\right]_{\sigma_1}^{\sigma_2\sigma_{X}}F^{(34)}_X\right.\notag\\
		&~~~\times \left.\left[{\cal A}_{L_2,S_2}\right]_{\sigma_{X}}^{\sigma_3\sigma_4}~\right),
		\label{eq:three_body_decay}
	\end{align}
	where $F^{(34)}_X$ is the line shape of the intermediate particle-$X$, which can be Breit-Wigner or any other formalism; 
	$\cal{A}$ can be either $\cal{H}$ or $\cal{C}$, corresponding to PWF-H or PWF-C.
	
	\section{PWF with identical particles}
	\label{sec:identical}
	
	When the final state contains identical particles, the amplitude is symmetric (for identical bosons) or antisymmetric (for identical fermions) under identical particle permutations. 
	For 2BD with identical final state particles, the PWF satisfies
	\begin{equation}
		\left[{\cal A}_{L,S}\right]_{\sigma_1}^{\sigma_2\sigma_3}=\pm~\left[{\cal A}_{L,S}\right]_{\sigma_1}^{\sigma_3\sigma_2},
	\end{equation}
	where $+$ and $-$ are for the symmetric and antisymmetric, respectively.
	By employing Eq.~\eqref{eq:PWF_A_cmf}, one can obtain the following selection rule: $$(-1)^{L+S}=1.$$
	Only $L$-$S$ combinations satisfied this condition are theoretically allowed. 
	This result means that the presence of identical particles reduces the allowed number of $L$-$S$ combinations in PWF.
	
	For multi-body decay with identical particles, the situation will become complicated. 
	For instance, we can reconsider the 3-body decay as discussed in Sec.~\ref{sec:manybodydecay}. 
	If particle-3 and particle-4 are identical particles, the case is similar as that in 2BD. 
	Since one only needs to consider the permutation symmetry of the PWF of the process $X\to 3+4$ as shown in Eq.~\eqref{eq:three_body_decay}. 
	If particle-2 and particle-3 are identical particles, the case will be different. 
	By performing identical particle permutation on the PWF in Eq.~\eqref{eq:three_body_decay}, one can obtain the following PWF:
	\begin{align}
		\left[{\tilde{\cal A}}_{L_1,S_1,L_2,S_2}\right]_{\sigma_1}^{\sigma_2\sigma_3\sigma_4}=\sum_{\sigma_X}&\left(~\left[{\cal A}_{L_1,S_1}\right]_{\sigma_1}^{\sigma_3\sigma_{X}}F^{(24)}_X\right.\notag\\
		&~~~\times \left.\left[{\cal A}_{L_2,S_2}\right]_{\sigma_{X}}^{\sigma_2\sigma_4}~\right).
		\label{eq:three_body_decay_permutation}
	\end{align}

	The meanings of $L_i$ and $S_i$ $(i=1,2)$ in the $\mathcal{A}$ from Eq.~\eqref{eq:three_body_decay} and in the $\tilde{\mathcal{A}}$ from Eq.~\eqref{eq:three_body_decay_permutation} are different, as they represent the orbital-spin quantum numbers between different subsystems of the final state. 
	Therefore, there cannot be only a sign difference between $\mathcal{A}$ and $\tilde{\mathcal{A}}$.\footnote{Essentially, this is because the complete set of commuting operators (CSCO) formed by the angular momentum operators is not closed under permutations of identical particles. 
		Strictly speaking, when dealing with processes involving identical particles, one needs to construct a closed CSCO under permutations, and select the common eigenstates of the operators in the CSCO as the orthogonal and complete basis for amplitude analysis.} 
	However, we can simply define permutation amplitudes (PAs) as follows:
	\begin{equation}
		\left[{\cal P\hspace{-0.6mm}A}_{L_1,S_1,L_2,S_2}\right]=\left[{\cal A}_{L_1,S_1,L_2,S_2}\pm\tilde{{\cal A}}_{L_1,S_1,L_2,S_2}\right],
		\label{eq:PWF_permutation}
	\end{equation}
	where we have omitted the polarization indices $\sigma_i~(i=1,2,3,4)$, and $+$($-$) is for the identical bosons (fermions).
	PAs automatically satisfy the permutation symmetry of identical particles. 
	It should be noted that PAs do not have definite orbital-spin quantum numbers, so there may be overlap between PAs labeled by different $L_i$-$S_i~(i=1,2)$ combinations. 
	Nevertheless, using PAs as the PWFs for processes involving identical particles is very convenient.

	\section{Summary}
	\label{sec:summary}
	
	In this work, we propose a concise algorithm for computing PWF of arbitrary processes under the covariant $L$-$S$ scheme, thereby enabling automated calculations under this scheme. 
	It is worth mentioning that, within this framework, we also provide PWF under the helicity scheme. 
	Furthermore, the framework is applicable to scenarios involving both massive and massless particles. 
	Additionally, we discuss PWF for cascade decays and processes involving identical particles. 
	To facilitate a better understanding of the algorithm proposed here, we have developed a package\footnote{For more details, please refer to the website: https://github.com/Wu-ShuMing/PWFs.} based on C++. 
	This package can automatically calculates PWF for processes of 2BD under both the helicity scheme and the covariant $L$-$S$ scheme in any reference frame. 
	It is hoped that this work will provide researchers with more options for PWA, aiming to conduct more comprehensive and systematic analysis of experimental data.

	\begin{acknowledgments}
		We thank useful discussions with Yi Jiang, Run-Qiu Ma, Shi Wang, Wen-Bin Qian, Bei-Jiang Liu, Xiao-Rui Lyu and Shuang-Shi Fang.
		This work is partly supported by the China Postdoctoral Science Foundation under Grants No. 119103S408 (H.J.J.),
		and by National Natural Science Foundation of China under Grants No.12175239, 12221005 (J.J.W.),
		and by the National Key Research and Development Program of China under Contract No. 2020YFA0406400 (J.J.W.), 
		and also by Chinese Academy of Sciences under Grant No. YSBR-101 (J.J.W.), 
		and also by Xiaomi Foundation / Xiaomi Young Talents Program (J.J.W.).
		
	\end{acknowledgments}
	
	\bibliography{refs.bib}
	\bibliographystyle{unsrt}
	
	\begin{appendix}
		
		\section{The dilation scale}
		\label{appd:scale}
		The explicit form of the dilation scale is as follows: 
		\begin{widetext}
			\begin{align}
				I\left(|\mathbf{p}|,s,\sigma\right)=
				\left\{\begin{array}{l}
					\sum_{k=\max{\left(\sigma-\frac{s}{2},-\frac{s}{2}\right)}}^{\min{\left(\sigma+\frac{s}{2},\frac{s}{2}\right)}}~\frac{\sqrt{\pi} (2 s+1) (s-\sigma)! (s+\sigma)!s !~ \vartheta^{2 k-\sigma}}{2^{(2 s+1)} \left(s+\frac{1}{2}\right) !\left(\frac{s}{2}-k\right) !\left(\frac{s}{2}+k\right) !\left(\frac{s}{2}+k-\sigma\right) !\left(\frac{s}{2}-k+\sigma\right) !},\text{(for integer spin $s$)}
					\\
					\sum_{k=\max{\left(\sigma-\frac{2s-1}{4},-\frac{2s+1}{4}\right)}}^{\min{\left(\sigma+\frac{2s-1}{4},\frac{2s+1}{4}\right)}}~\frac{\sqrt{\pi} (2 s+1) (s-\sigma)! (s+\sigma)! \left(s-\frac{1}{2}\right) !~\left(\vartheta^{2 k-\sigma}-\vartheta^{\sigma-2 k} \right)}{2^{(2 s+1)} s !\left(\frac{s}{2}-k+\frac{1}{4}\right) !\left(\frac{s}{2}+k+\frac{1}{4}\right) !\left(\frac{s}{2}+k-\sigma-\frac{1}{4}\right) !\left(\frac{s}{2}-k+\sigma-\frac{1}{4}\right) !}, \text{(for half-integer spin $s$)}
				\end{array}\right.
			\end{align}
		\end{widetext}
		with
		\begin{equation}
			\vartheta_i=\left\{\begin{array}{cl}
				(|\mathbf{p}_i|+E_i)/m_i & \text{~~~for massive particle}\\
				|\mathbf{p}_i|/|\mathbf{k}_i| & \text{~~~for massless particle}
			\end{array}\right.,\notag
			\vspace{3mm}
			\label{eq:covariant_factor}
		\end{equation}
		where $m_i$ and $E_i=\sqrt{m_i^2+|\mathbf{p}_i|^2}$ is the mass and energy of the particle-$i$, respectively; $\mathbf{k}_i$ is the standard momentum of the massless particle, which can be freely selected.

	\end{appendix}
	
\end{document}